\def\year{2021}\relax
\documentclass[letterpaper]{article} % DO NOT CHANGE THIS
\usepackage{aaai21}  % DO NOT CHANGE THIS
\usepackage{times}  % DO NOT CHANGE THIS
\usepackage{helvet} % DO NOT CHANGE THIS
\usepackage{courier}  % DO NOT CHANGE THIS
\usepackage[hyphens]{url}  % DO NOT CHANGE THIS
\usepackage{graphicx} % DO NOT CHANGE THIS
\usepackage{natbib}  % DO NOT CHANGE THIS AND DO NOT ADD ANY OPTIONS TO IT
\usepackage{caption} % DO NOT CHANGE THIS AND DO NOT ADD ANY OPTIONS TO IT
\usepackage{amsmath}
\usepackage{booktabs}
\usepackage{algorithm}
\usepackage{amsfonts}
\usepackage{comment}
\usepackage{color}

\title{Similarity Metrics for Transfer Learning in Financial Markets}
\author{
    Diego Pino$^1$,
    Javier García$^1$,
    Fernando Fernández$^1$,
    Svitlana S Vyetrenko$^2$
    \\
}
\affiliations{
    $^1$Departamento de Inform\'atica, Universidad Carlos III de Madrid\\
    
    Avda. de la Universidad, 30. 28911 Legan\'es (Madrid). Spain\\
    $^2$J. P. Morgan AI Research, New York, NY, USA\\
    \{dpino,fjgpolo,ffernand\}@inf.uc3m.es, svitlana.s.vyetrenko@jpmchase.com
}

\begin{document}

\maketitle

\begin{abstract}

Markov Decision Processes (MDPs) are an effective way to formally describe many Machine Learning problems. In fact, recently MDPs have also emerged as a powerful framework to model financial trading tasks. For example, financial MDPs can model different market scenarios. However, the learning of a (near-)optimal policy for each of these financial MDPs can be a very time-consuming process, especially when nothing is known about the policy to begin with. An alternative approach is to find a similar financial MDP for which we have already learned its policy, and then reuse such policy in the learning of a new policy for a new financial MDP. Such a knowledge transfer between market scenarios raises several issues. On the one hand, how to measure the similarity between financial MDPs. On the other hand, how to use this similarity measurement to effectively transfer the knowledge between financial MDPs. This paper addresses both of these issues. Regarding the first one, this paper analyzes the use of three similarity metrics based on \textit{conceptual}, \textit{structural} and \textit{performance} aspects of the financial MDPs. Regarding the second one, this paper uses Probabilistic Policy Reuse to balance the exploitation/exploration in the learning of a new financial MDP according to the similarity of the previous financial MDPs whose knowledge is reused.

%Transfer in Reinforcement Learning (RL) aims to remedy the problem of learning complex RL tasks from scratch, which is impractical in most of the cases due to the huge sample requirements. To overcome this problem, transferring the knowledge acquired from a set of source tasks to a new target task is a core idea. This technique is used in this work in the field of financial markets, with the objective of transferring the knowledge acquired in simulated scenarios to the real one. The first step is to calculate the distance between real and simulated scenarios, in order to determine how good the knowledge transfer will be. This distance is calculated through 3 different approaches based on pi-reuse, Restricted Boltzmann Machines and stylized facts. Finally PRQ-Learning is used to train an agent and is compared with Q-Learning achieving better results.
\end{abstract}

\section{Introduction}

Markov decision processes (MDPs) are a common way of formulating decision making problems in reinforcement learning (RL) tasks~\cite{sutton2018reinforcement}. An MDP provides a standard  formalism for multi-stage decision processes, whilst at the same time being able to capture the stochastic nature of realistic situations. This is the reason why MDPs have also emerged as a powerful framework for modeling real financial trading problems~\cite{chakraborty2019capturing,huang2018financial,bauerle2011markov}. In these financial MDPs, the objective is to learn a trading strategy or policy able to maximize some measure of performance over time, typically the profit. Interestingly, these financial MDPs can be configured to model different market scenarios, so that different trading policies can be tested and analyzed. However, finding these policies from scratch is often a hard task. On the one hand, the learning process is based on a computationally intensive trial-and-error process guided by reward signals from the environment. On the other hand, such a trial-and-error process is unfeasible in a real trading scenario where a single bad decision can lead to large losses. Therefore, when it is possible to obtain these trading policies, it would be good to take advantage of them as much as possible. Specifically, rather than learning a new trading policy for every financial MDP, a policy could be learned on one financial MDP, then transferred to another, similar financial MDP, and either used as is, or treated as a starting point from which to learn a new trading policy. 

Such a knowledge transfer is particularly interesting in a \textit{Sim-to-Real} scenario~\cite{tan2018sim}. Due to the reality gap, trading strategies learned in simulation usually do not perform well in real environments. Since the reality gap is caused by model discrepancies between the simulation and the real dynamics, similarity metrics can be used to effectively quantify that gap, and as a direct way to measure simulation fidelity. In this way, the hedge funds, investors, or banks can test their trading strategies in simulation before risking their funds in a real trading problem~\cite{vyetrenko2019real}. In this context, similarity metrics can be used to select the best simulated financial MDP to transfer from, avoiding real losses.

Therefore, it is a critical step to decide when two MDPs are similar. This paper investigates three different strategies to compute the similarity between MDPs in a financial context. These strategies allows to measure the similarity from three different perspectives considering \textit{conceptual}, \textit{structural} and \textit{performance} aspects of the MDPs. The \textit{conceptual} similarity is based on comparing the statistical properties (also known as \textit{stylized facts}~\cite{vyetrenko2019real}) of the asset returns, order volumes, order arrival times, order cancellations, etc. Typically, such stylized facts are used by human experts to analyze the similarities and differences between simulated and real financial data~\cite{fabretti2013problem}. In contrast, the \textit{structural} similarity is based on the comparison of the experience tuples generated by two different MDPs: two MDPs are considered to be similar whenever the experience tuples they generate are similar as well. In this paper, such a structural comparison is performed by using restricted Bolztmann machines (RBMs)~\cite{inproceedings}. Finally, this paper uses the well-known exploitation/exploitation strategy, $\pi$-reuse, to measure the \textit{performance} similarity between MDPs. In particular, it measures the reuse gain (i.e., the ``advantage'') of using one task to aid the learning of another~\cite{Fernandez2013}. The higher the reuse gain, the greater the similarity between the MDPs. Finally, such a reuse gains are used by a Policy Reuse algorithm to bias the learning process of a new MDP in a transfer learning context.

This paper is organized as follows. Section \textit{Related Work} describes some previous related work. Section \textit{Background} presents key concept on RL, transfer learning and similarity metrics required to better understand the rest of the paper. Section \textit{Abides} introduces the abides simulator used to recreate different financial markets. Section \textit{Similarity Metrics} presents the similarity metrics based on the conceptual, structural and performance relationships between the different financial markets. Finally, Section \textit{Experimentation} shows the evaluation performed, and Section \textit{Conclusions} summarizes the conclusions and some future work.

%However, sometimes the cost of such learning process may be very high. For that reason, exists techniques which reuse the knowledge acquired in past tasks to reduce the cost of the current learning process. The similarity between the past tasks and the new one, determine how good will be the transfer of knowledge. Therefore, finding a similarity metric, that gives us the distance between tasks correctly, is very important.  

%In this paper, transfer learning will be used to reuse the knowledge learning in 6 simulated scenarios to reduce the cost of the real scenario learning process, using Abides to simulate the markets. In addition, Three similarity metrics will be studied to obtain the distance between scenarios and to determine which of them are more similar to each other and why. Such metrics are based in three different approaches. The first one, is based in the reuse gain obtained applying $\pi$-reuse. The second one, used Restricted Boltzmann Machines to determine the distance between tasks. And the last one, is based on the structure of the financial markets themselves, called stylized facts.

\section{Related Work}
\label{sec:relatedwork}

RL has been used in several applications in finance and trading, including portfolio optimization and optimal trade execution. Given that actions taken in finance and trading may have long-term effects not immediately measurable, some financial problems can be viewed as sequential decision problems. Furthermore, the environments in which these areas work may be very large or continuous, so RL can well-suited to solving finance problems. RL was first introduced and implemented in the financial market in 1997~\cite{Performancefunctions}.

Some examples where RL has been successfully applied to financial problems include portfolio optimization, optimized trade execution, and market-making. In portfolio optimization the goal is to create an optimum portfolio given the specific factors that should be maximized or minimized (e.g. expected return, financial risk) and taking into account any  constraints~\cite{kanwar2019deep}. Instead, the goal of optimized trade execution is to sell or buy a specific number of shares of a stock in a fixed time period, such that the revenue received (in the case of selling) is maximized or the capital spent (in the case of buying) is minimized~\cite{nevmyvaka2006reinforcement}. In market-making, the market maker buys and sells stocks with the goal of maximizing the profit from buying and selling them and minimizing the inventory risk. In this context, RL has been used successfully to come up with price setting strategies to maximize profit and minimize inventory risk~\cite{ganesh2019reinforcement}. However, in contrast to these approaches, this work focuses on a different objective: to measure the similarity of financial markets, modeled in the multi-agent simulator Abides, in order to determine which scenarios are most similar to the real world and transfer the knowledge learned to the real world to maximize profit. Transfer RL has also been used in a financial context~\cite{martinez2020probabilistic} using PPR to identify the similarity among different market scenarios in the context of pricing. Identifying these similarities permit to measure how a market scenario is similar to another, and thus how the pricing policy learned in a market scenario is useful to learn a pricing policy in a related, but different, market scenario.

\section{Background} 
\label{sec:background}

This section introduces key concepts required to better understand the rest of the paper. First, some background in RL is introduced, then the main concepts of transfer RL are visited, and finally the concepts of similarity and distance are described.

\subsection{Reinforcement Learning}
\label{subsec:backrl}

RL~\cite{Kaelbling1996ReinforcementLA} is an area of machine learning where agents learn what actions to take in an environment to maximize an accumulated reward. The learning process is based on trial and error guided by reinforcement signals from the environment that reward agents for performing actions that bring them closer to solving the problem. A combination of exploration (trying the unknown) and exploitation (using knowledge the agent already has) can be used to make improvements in the performance of RL algorithms.

In particular, RL tasks are described as Markov Decision Processes (MDPs) represented by tuples in the form $\mathcal M=\langle S, A, T, R \rangle$, where $S$ is the state space, $A$  is the action space, $T : S \times A \rightarrow S$ is the transition function between states, and $R : S \times A \rightarrow \mathbb{R}$ is the reward function. At each step, the agent is able to observe the current state, and choose an action according to its policy $\pi : S \rightarrow A$. The goal of the RL agent is to learn an optimal policy $\pi^*$ that maximizes the return $J(\pi)$:
\begin{equation}
    J(\pi) = \sum_{k=0}^{K}\gamma^k r_k
\label{eq:sumrewards}
\end{equation}
\noindent
where $r_k$ is the immediate reward obtained by the agent on step $k$, and $\gamma$ is the discount factor, which determines how relevant the future is (with $0 \leq \gamma \leq 1$). The interaction between the agent and the environment tends to be broken into episodes, that end when reaching a terminal state, or when a fixed amount of time has passed. With the goal of learning the policy $\pi$, Temporal Differences methods~\cite{sutton2018reinforcement} estimate the sum of rewards represented in Equation~\ref{eq:sumrewards}. The function that estimates the sum of rewards, i.e., the return for each state $s$ given the policy $\pi$ is called the value-function $V^{\pi}(s) = E[J(\pi)|s_0 = s]$.  Similarly, the action-value function $Q^{\pi}(s,a) = E[J(\pi)|s_0 = s, a_0=a]$ is the estimation of the value of performing a given action $a$ at a state $s$ being $\pi$ the policy followed. The Q-learning algorithm~\cite{watkins1989} is one of the most widely used for computing the action-value function.

\subsection{Transfer Learning}
\label{subsec:backtransferrl}

The learning process in RL can sometimes be too costly, thus the concept of transfer learning was born. In the transfer learning scenario we assume there is an agent who previously has addressed a set of source tasks represented as a sequence of MDPs, $\mathcal M_{1},\dots, \mathcal M_{n}$. If these tasks are somehow ``similar'' to a new task $\mathcal M_{n+1}$, then it seems reasonable the agent uses the acquired knowledge solving $\mathcal M_{1},\dots, \mathcal M_{n}$ to solve the new task $\mathcal M_{n+1}$ faster than it would be able to from scratch. Transfer learning is the problem of how to obtain, represent and, ultimately, use the previous knowledge of an agent~\cite{torrey2010transfer,taylor2009transfer}.

However, transferring knowledge is not an easy endeavour. On the one hand, we can distinguish different transfer settings depending on whether the source and target tasks share or not the state and action spaces, the transition probabilities and the reward functions. It is common to assume that the tasks share the state space and the action set, but differing the transition probabilities and/or reward functions. However, in case the tasks do not share the state and/or the action spaces, it is required to build mapping functions, $\mathcal X_{S}(s_{t})=s_{s}$, $\mathcal X_{A}(a_{t})=a_{s}$, able to map a state $s_{t}$ or action $a_{t}$ in the target task to a state $s_{s}$ or action $a_{s}$ in the source task. Such mapping functions require not only knowing if two tasks are related, but \textit{how} they are related, which means an added difficulty. On the other hand, it is required to select what type of information is going to be transferred. Different types of information have been transferred so far ranging from instance transfer (a set of samples collected in the source task) to policy transfer (i.e., the policy $\pi$ learned in the source task). Nor is this a simple task, because depending on how much and how the source and target tasks are related, it could be transferred one type of information or another. For example, some of them transfer the whole Q-value function~\cite{MahmudM.M.Hassan2013CMDP}, others transfer the learned policy~\cite{Mehta2008}, or they could transfer a model of the state transition function and/or the reward function~\cite{Sunmola2006}. But this paper focuses on reusing the policy by $\pi$-reuse.

Finally, the most ``similar'' task among $\mathcal M_{1},\dots, \mathcal M_{n}$ to solve $\mathcal M_{n+1}$ should be selected in the hope that it produces the most positive transfer. For this purpose, similarity metrics could be used, which translate into a measurable quantity of how related two tasks are.

\subsection{Similarity Metrics}
\label{subsec:simdist}

Similarity metrics are a very important part of transfer learning, as they provide a measure of distance between tasks. Similarity functions, or their complementary distance functions, are mathematical functions that assign a numerical value to each pair of concepts or objects in a given domain. This value measures how similar these two concepts or objects are: if they are very similar, it is assigned a very low distance, and if they are very dissimilar, it is assigned a larger distance~\cite{ontanon2020overview}. Therefore, through similarity metrics it is possible to determine how good the transfer of knowledge will be.

\subsection{Simulated Environment}
\subsubsection{Limit Order Books}
The limit order book represents a snapshot of the supply and demand for an asset at a given point in time. It is an electronic record of all the outstanding buy and sell limit orders organized by price levels. A matching engine, such as first-in-first-out (FIFO), is used to pair incoming buy and sell order interest \cite{Bouchaud_book}. The limit order book is split into two sides--the ask and bid sides containing all the sell and buy limit orders, respectively. At time $t$, let $b_t$ be the best bid price, and let $a_t$ be the best ask price. We define mid-price as $m_t =\frac{a_t+b_t}{2}$. Choose time scale $\Delta t$. Given a time scale $\Delta t$, which can range from milliseconds to months, the log return (or simply return) at scale $\Delta t$ is defined as $r_{t, \Delta t} = \ln m_{t+\Delta t} - \ln m_t$.

%The difference between the lowest ask price (best ask) and highest bid price (best bid) is called the {\it spread}. The {\it mid-price} is the average of the best bid and ask prices. The {\it depth of placement} is the number of levels on which orders are placed into the order book counting from the mid-price.

\begin{figure}[!htb]
 \centering
 \includegraphics[width=0.35\textwidth]{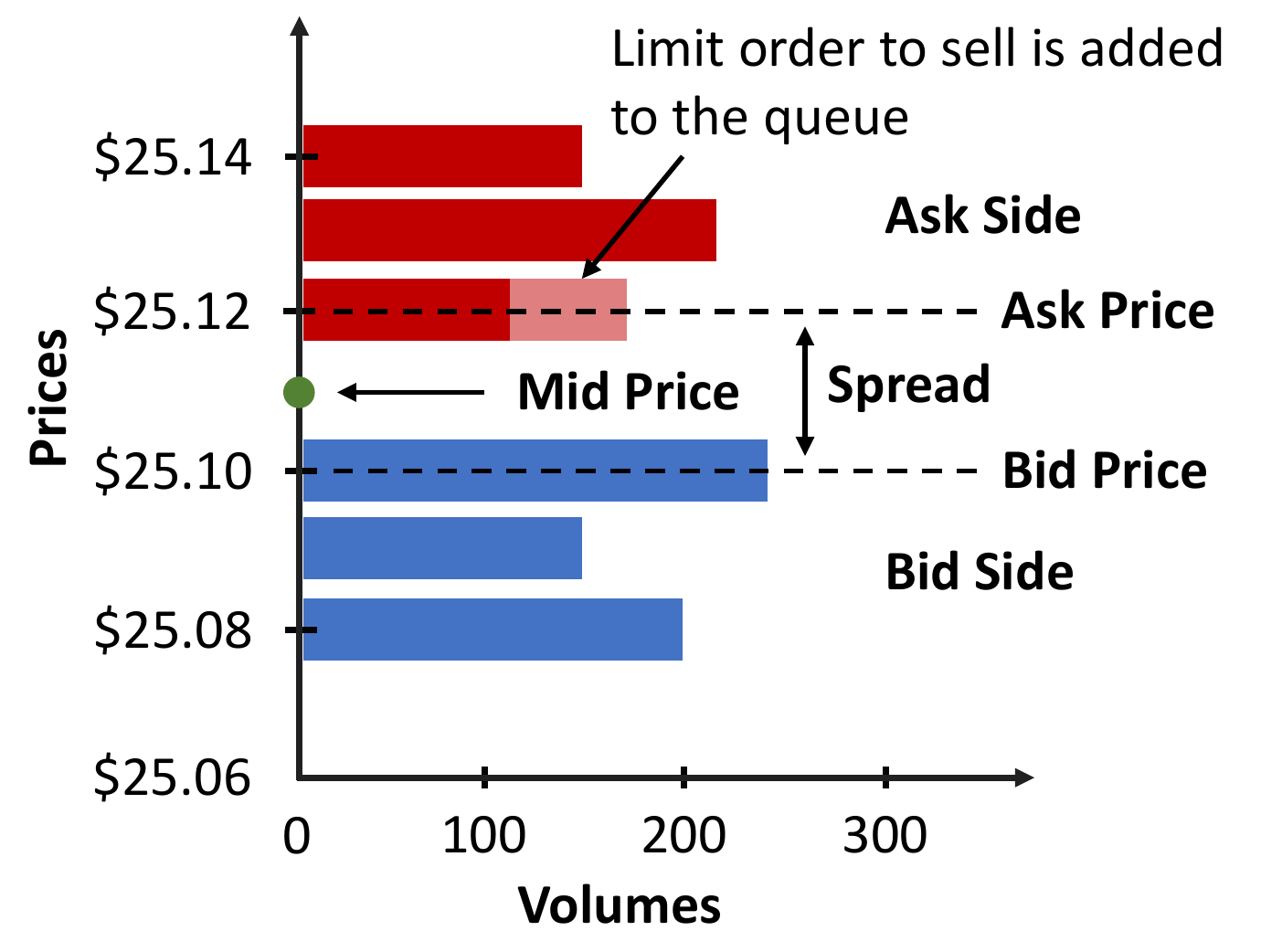}
 \caption{Visualization of the limit order book structure.}
 \label{fig:LOB}
\end{figure}

Order types are further distinguished between limit orders and market orders. A limit order specifies a price that should not be exceeded in the case of a buy order, or should not be gone below in the case of a sell order. Hence, a limit order queues a resting order in the order book at the side of the market participant. A market order indicates that the trader is willing to accept the best price available immediately. A diagram illustrating the limit order book structure is provided in Figure~\ref{fig:LOB}.

\subsubsection{ABIDES Simulator}
\label{sec:abides}
ABIDES~\cite{byrd2019abides}  -- which stands for  Agent Based Interactive Discrete Event Simulator -- was designed from the ground up to support AI agent research in market applications, and currently enables the simulation of tens of thousands of trading agents interacting with an exchange agent to facilitate transactions. ABIDES supports configurable pairwise network latencies between each individual agent as well as the exchange. Similar to the real exchange, ABIDES allows to simulate market activity as a result of interaction of multiple background agents with diverse objectives.

We use ABIDES to simulate market scenarios which are distinguished by different number of background agents. Each scenario has a unique agent configuration making each learning process a different task. In particular, we use configurations with {\bf{non-learning}} exchange, zero intelligence, momentum and noise traders as well a {\bf{learning}} $Q$-Learner agent. 

The $Q$-learner agent is a basic agent that will try to learn to beat the market by using reinforcement learning. To do this, at each instant of time, it will have to perform one of the following actions: buy an asset, sell an asset or do nothing. The agent will not be able to buy another asset if he already has one, just as he will not be able to sell one if he already owes one. The agent's state is represented by the difference between the buying and selling volume, and the information whether he currently owes, owns or doesn't own an asset. 

A brief description of non-learning background agents is given below.

{\bf{Exchange agent}:}
NASDAQ-like agent which lists any number of securities for trade against a LOB with price-then-FIFO matching rules, and a simulation kernel which manages the flow of time and handles all inter-agent communication. 

{\bf Zero intelligence agents:} 
In our experiment, we use the implementation of zero intelligence agents described in \cite{byrd2019explaining} to represent institutional investors who follow mean reverting fundamental price series to provide liquidity to the market via limit orders.

{\bf Momentum agents:} The momentum agents base their trading decision on observed price trends. Our implementation compares the 20 past mid-price observations with the 50 past observations and places a buy order of random size, if the former exceeds the latter and a sell order otherwise. 

{\bf Noise agents:} Noise agents are designed to emulate the actions of consumer agents who trade on demand without any other considerations (e.g., \cite{kyle1985continuous}).   Each retail agent trades once a day by placing a market order. The direction and the size of the trade are chosen randomly. 

\section{Similarity Metrics}
\label{sec:similaritymetrics}
This paper uses three different concept of metrics which are based on \textit{conceptual}, \textit{structural} and \textit{performance} aspects of the MDPs to be compared. The first one, called stylized facts, is based on the features of the financial markets themselves. The second one uses Restricted Boltzmann Machines (RBM) to determine the structural similarity between tasks. Finally, the last one is based in the reuse gain obtained applying $\pi$-reuse. 

\subsection{Stylized Facts}

The conceptual metric is based on the use of what it is known as stylized facts~\cite{vyetrenko2019real,RCont2001}. The result of more than half a century of empirical studies on financial time series indicates that although different assets are not necessarily influenced by the same events or information sets, if one examines their properties from a statistical point of view, the seemingly random variations of asset prices share some quite nontrivial statistical properties. Such properties, common across a wide range of instruments, markets and time periods are called stylized empirical facts. The stylized facts are not a metric per se, but are actually based on an analysis of the graphs by studying the shape of the graphs.
Vyetrenko et al. (\citeyear{vyetrenko2019real}) describe several groups of stylized facts, but this paper uses only those stylized facts related to asset return distributions. In particular, we focus on:

\begin{itemize}
    \item {\bf Mid Price Returns}: Statistical similarity of asset mid price returns $r_{t, \Delta t}$ for $\Delta t = \text{1}$ and $\Delta t =\text{10}$ minutes.
    \item {\bf Auto-correlation}: Statistical similarity of linear auto-correlations of asset mid price returns $\text{corr}(r_{t+\tau, \Delta t},r_{t, \Delta t})$  over $\Delta t = \text{30}$ minutes. 
\end{itemize}

\subsection{Restricted Boltzmann Machines}

RBMs~\cite{inproceedings} are energy-based models for unsupervised learning. They use a generative model of the distribution of training data for prediction. These models employ stochastic nodes and layers, making them less vulnerable to local minima~\cite{RBMS}. RBMs are stochastic neural networks with bidirectional connections between the visible and hidden layers. This allows RBMs to posses the capability of regenerating visible layer values, given a hidden layer configuration. The visible layer represents input data, while the hidden layer discovers more informative spaces to describe input instances. Therefore, RBMs could also be seen as density estimators, where the hidden layer approximates a (factored) density representation of the input units. Formally, an RBM consists of two binary layers: one visible and one hidden. The visible layer models the data while the hidden layer enlarges the class of distributions that can be represented to an arbitrary complexity. 

The main idea is that we can use an RBM to describe different MDPs in a common representation, providing a similarity measure. The RBMs are good as similarity metrics since they can automatically discover the dynamic phases of MDPs, and predict the transfer performance between source and target tasks. Moreover, they have been used in many domains for this purpose with good results~\cite{inproceedings}.

We first train an RBM to model data collected in the source task, yielding a set of relevant and informative features that characterize the source MDP. These features can then be tested on the target task to assess MDP similarity. To do this, we collect the tuples from the source task and pass them as input to the RBM, the expected output being the input tuples, the weights of the neurons are adjusted by contrastive divergence. Then, the tuples of the target task are collected and provided to the RBM, already trained, and we calculate the mean square error obtained between the input and the output, so that the lower the error the higher the similarity.

\subsection{$\pi$-reuse}
\label{sec:pireuse}

The $\pi$-reuse strategy is an exploration strategy able to bias a new learning process with a past policy~\cite{Fernandez2013}. Let $\Pi_{past}$ be the past policy to reuse and $\Pi_{new}$ the new policy to be learned. We assume that we are using a RL algorithm to learn the action policy, so we are learning the related $Q$ function. Any RL algorithm can be used to learn the $Q$ function, with the only requirement that it can learn off-policy, i.e., it can learn a policy while executing a different one, as Q-Learning does.

The goal of $\pi$-reuse is to balance random exploration, exploitation of the past policy, and exploitation of the new policy, as represented in Equation~\ref{eq:pi-reuse}.

\begin{equation}
a= \left\{\begin{array}{ll}
\Pi_{past}(s) & \textrm{w.p. } \psi \\
\epsilon-greedy(\Pi_{new}(s)) & \textrm{w.p. } (1-\psi) \\
\end{array}  \right.
\label{eq:pi-reuse}
\end{equation}

The $\pi$-reuse strategy follows the past policy with probability $\psi$, and it exploits the new policy with probability of $1-\psi$. As random exploration is always required, it exploits the new policy with a $\epsilon$-greedy strategy.

Given a policy $\Pi_{past}$ that solves a task $\Omega_{past}$, and a new task $\Omega$, the reuse Gain of the policy $\Pi_{past}$ on the task $\Omega$, the Reuse Gain of the policy $\Pi_{past}$ on the task $\Omega$, is the gain obtained when applying the $\pi$-reuse exploration strategy with the policy $\Pi_{past}$ to learn the policy $\Pi_{new}$. The Reuse Gain is used in this work to measure the distance between the scenarios, taking into account that the higher the reuse gain, the lower the distance.

\section{PRQ-Learning}

PRQ-Learning is an algorithm based on $\pi$-reuse and its goal is to solve a new task reusing the knowledge in a library of pasts tasks~\cite{Fernandez2013}. Therefore, PRQ-Learning is equipped with a policy library composed of $n$ past optimal policies that solve $n$ different tasks, respectively, plus the ongoing learned policy. PRQ-Learning is able to select which policy should be reused and what exploration/exploitation strategy follow. 

Let $W_{i}$ be the Reuse Gain of the policy $\Pi_{i}$ on the task $\Omega$. Also, let $W_{\Omega}$ be the average reward that is received when following the policy $\Pi_{\Omega}$ greedily. The solution we introduce consists of following a softmax strategy using the values $W_{\Omega}$ and $W_{i}$, with a temperature parameter $\tau$, as shown in Equation~\ref{eq:selectprob}. This value is also computed for $\Pi_{0}$, which we assume to be $\Pi_{\Omega}$.
%Equation~\ref{eq:selectprob} provides a way of deciding the policy $\Pi_{k}$ to reuse at each learning episode.

\begin{equation}
\begin{split}
%\Pi_{k} = &\arg_{\Pi_{j}} max(P(\Pi_{j})), \\  where
\ P(\Pi_{j}) 
= &\frac{e^{\tau W_{j}}}{
\sum_{p=0}^{n} e^{\tau W_{p}}
}
\end{split}
\label{eq:selectprob}
\end{equation}

In the first episode, all the policies have the same probability to be chosen. Once a policy is chosen, the algorithm reuse it to solve the task, updating its reuse gain with the reward obtained in the episode, and therefore, updating the probability to follow each policy. The policy being learned can also be chosen, although in the initial steps it behaves as a random policy, given that the $Q$ values are initialized to $0$. While new updates are performed over the $Q$-function, it becomes more accurate, and receives higher rewards when executed. After executing several episodes, it is expected that the new on-going policy obtains higher gains than reusing the past policies, so it will be chosen most of the time. 

If the policy being learned is chosen, the algorithm follows a completely greedy strategy. However, if the policy chosen is one of the past policies, the $\pi$-reuse strategy, is followed instead. In this way, the reuse gain of each of the past policies can be estimated online with the learning of the new policy. 

\section{Experimentation}
\label{sec:experimentation}

This section evaluates the similarity metrics described in Section \textit{Similarity Metrics}. For this purpose, we will use $7$ different market scenarios, each one recreating different market conditions. Finally, PRQ-Learning is used to learn in a target market scenario, reusing the knowledge learned in previous source markets. However, the experimental setting is described first.

\subsection{Experimental setting}

This section first describes each of the 7 market scenarios used for the experimentation, and then it introduces the parameter setting used for each of the approaches.

\subsubsection{Market Scenarios}
\label{sec:marketscenarios}

There are 7 scenarios in this paper. The scenario 1 is the main scenario, i.e., we consider it as the target market scenario for the experiments with $\pi$-reuse and PRQ-Learning. Each scenario is composed of several agents that will determine the behavior of the system. The agents represents the traders who can place a buy or sell order. There are several types of agents in Abides, but in this paper we have used five, explained in the Section \textit{Abides}.

Furthermore, each of the market scenarios require a fundamental value, and at the end, the behaviour of all scenarios is determined by this value and the different agents that trade in the environment. Each market scenario is configured as described in Table~\ref{tab:agentconfiguration}, where the columns represent the scenario and the rows represent the number of agents of each type.

\begin{table}[htbp]
    \begin{center}
        \resizebox{0.49\textwidth}{!} {
            \begin{tabular}{ | c | c | c | c | c | c | c | c | }
                \hline & 1 & 2 & 3 & 4 & 5 & 6 & 7 \\ \hline
                Zero Intelligent & 100 & 100 & 100 & 100 & 100 & 100 & 100 \\ \hline
                Exchange & 1 & 1 & 1 & 1 & 1 & 1 & 1 \\ \hline
                Q-Learner & 1 & 1 & 1 & 1 & 1 & 1 & 1 \\ \hline
                Noise & 0 & 10 & 5 & 10 & 5 & 10 & 0 \\ \hline
                Momentum & 0 & 5 & 10 & 10 & 0 & 0 & 10 \\ \hline
            \end{tabular}
        }
    \end{center}
    \caption{Agent Configuration}
    \label{tab:agentconfiguration}
\end{table}

These agent configurations were chosen randomly by adding noise and momentum agents. The idea is to analyze which type of agents perturbs more the initial configuration and whether the similarity metrics determine that scenarios with less new agents are more similar to the initial configuration.

\subsubsection{Parameter Setting}
\label{sec:parametersetting}

The parameter setting for $Q$-learning is as follows. The parameter $\gamma$ is set to $0.98$, $\alpha$ to $0.99$, and it is decremented every step as $\alpha \times 0.999$. It uses $\epsilon$-greedy for the exploration/exploitation of the state and action space, where $\epsilon$ is set initially at $0.999$, and it is decremented every step as $\epsilon \times 0.9995$. Regarding $\pi$-reuse, $\psi$ is set to 1, and then it is reduced every step as $\psi \times 0.99$. Finally, in PRQ-Learning the temperature parameter $\tau$ is initially set to $0$ and it increases following a sigmoid function until $0.002$.

\begin{comment}
\begin{itemize}
    \item Q-Learning parameters:
        \begin{itemize}
          \item $\epsilon$: 0.999
          \item $\epsilon\_decay$: 0.9995
          \item $\alpha$: 0.99
          \item $\alpha\_decay$: 0.999
          \item $\gamma$: 0.98
          \item Exploration strategy: Epsilon greedy
          \item Learning episodes: Depending on the experiment
        \end{itemize}
        
    \item Pi-reuse parameters:
        \begin{itemize}
          \item $\psi$: 1
          \item $\psi\_decay$: 0.99
          \item $H$: Until the episode ends
          \item $K$: Depending on the experiment
        \end{itemize}
        
    \item PRQ-Learning parameters:
        \begin{itemize}
          \item $\psi$: 1
          \item $\psi\_decay$: 0.99
          \item Temperature: 0
          \item Temperature incremental: sigmoid growth until 0.002
          \item $H$: Until the episode ends
          \item $K$: Depending on the experiment
        \end{itemize}
\end{itemize}
\end{comment}

\subsection{Evaluation of the Similarity Metrics}
\label{subsec:evalmetrics}

This section presents the evaluation of the similarity between the different market scenarios by using the proposed styled facts, the RBMs and $\pi$-reuse.

\subsubsection{Stylized Facts}

We first compare distributional similarity between stylized facts. In the below analysis, scenarios that generate similar stylized fact distributions will be interpreted as having less distance between them. 

\begin{center}
    \begin{table}[htbp]
        \begin{tabular}{ c c } 
            \includegraphics[width=0.20\textwidth]{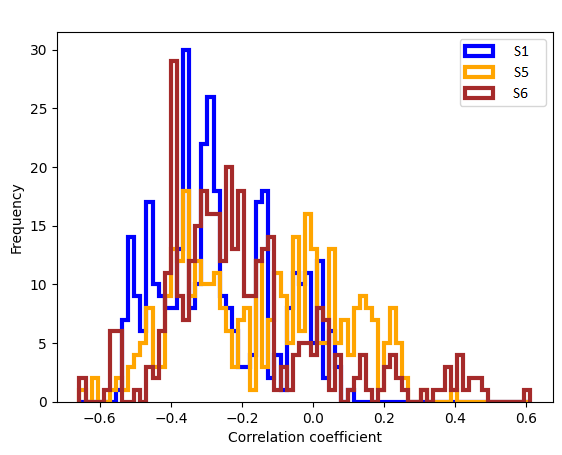} &
            \includegraphics[width=0.20\textwidth]{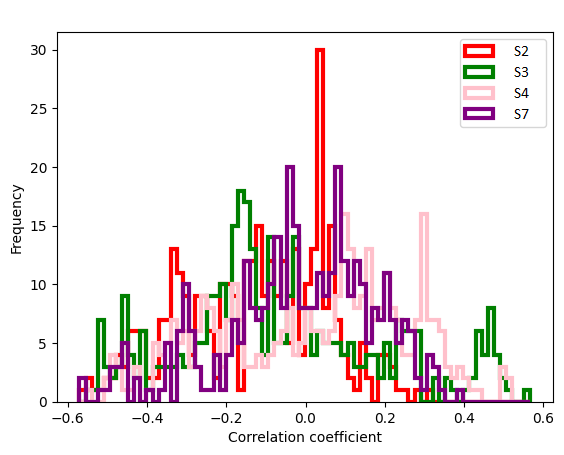} \\
            
            \includegraphics[width=0.20\textwidth]{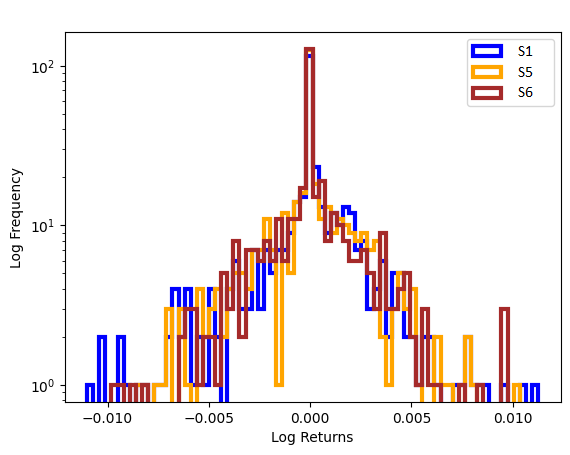} &
            \includegraphics[width=0.20\textwidth]{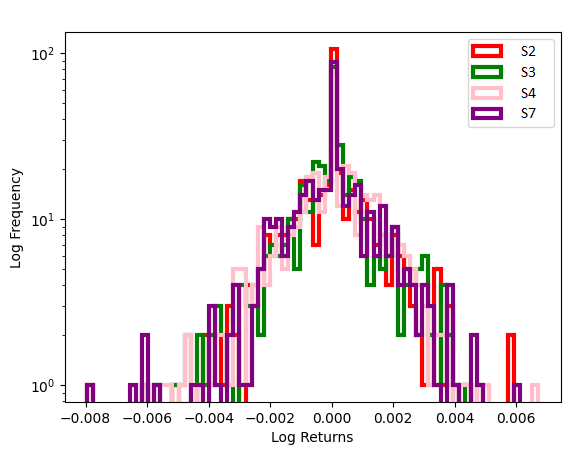} \\
        \end{tabular}
        \caption{Stylized Facts: Comparison between scenarios }
        \label{fig:StylizedFacts}
    \end{table}
\end{center}

Table~\ref{fig:StylizedFacts} shows the analysis of the stylized facts considered in this paper. The first row corresponds to autocorrelation and the second row corresponds to the minutely returns. Furthermore, the analysis of the stylized facts of the 7 proposed scenarios allows us to distinguish between two clusters: the first is composed of scenarios 1, 5, and 6, and the other of scenarios 2, 3, 4, and 7. For a better understanding, in Table~\ref{fig:StylizedFacts} the scenarios are already grouped according to the similarity in the shape of the stylized facts considered. Therefore, the graphs in the first column of Table~\ref{fig:StylizedFacts} show the value of the corresponding stylized fact to the markets 1, 5, and 6, whilst the graphs in the second column shows the stylized facts for the markets 2, 3, 4, and 7. The autocorrelation shows that in the cluster 1 the plot is skewed to the left, however in cluster 2 it is centered. Finally, the minutely returns in both clusters is similar but in the cluster 1 the graph is more wider.

\subsubsection{Restricted Boltzmann Machines}

\begin{figure}[htbp]
\includegraphics[width=0.5\textwidth]{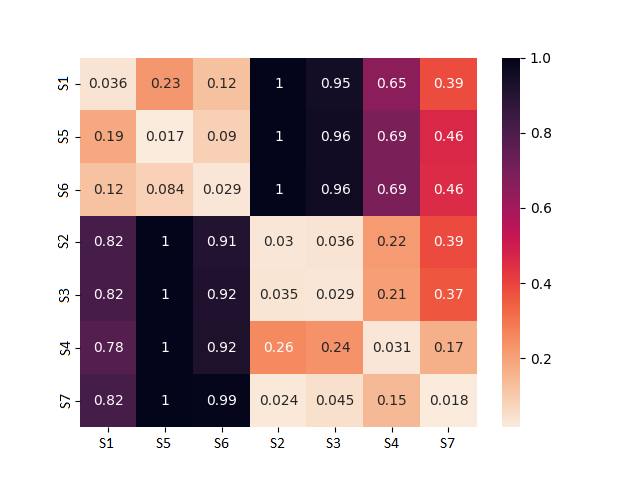}
\caption{Heat map with the distance between each pair of scenarios using Restricted Boltzmann Machines}
\label{fig:heatmap}
\end{figure}

The second metric analyzed in this paper is that provided by the RBMs. For this purpose, a set of transitions has been gathered in each of the proposed scenarios, and the distance between them has been computed as described in the section \textit{Restrictred Boltzmann Machines}. Figure~\ref{fig:heatmap} is a heat map that represents the distance calculated using RBMs from each pair of scenarios. In particular, the heat map shows as many rows and columns as there are market scenarios, and each cell represents the similarity of the X-axis market to the Y-axis market scenario. The distances range from 0 to 1, and a cell will be darker the greater the distance between the compared scenarios.
The order of the scenarios has been changed to facilitate the readability of the data.

In this case, two clusters are observed again. The first one indicates that the scenario 1, 5 and 6 are similar to each others, the second one indicates that the scenario 2, 3, 4 and 7 are also similar to each others. It is important to note that these results are the same as with the stylized facts and it is so interesting because scenario 1, 5 and 6 do not have momentum agents, and on the other hand, the scenarios 2, 3, 4 and 7 all have momentum agents. Somehow momentum agents have a critical influence in the functioning of the scenarios which allows us to separate the scenarios according whether they have momentum agents or not.

\subsubsection{$\pi$-reuse.}

\begin{figure}[htbp]
\includegraphics[width=0.5\textwidth]{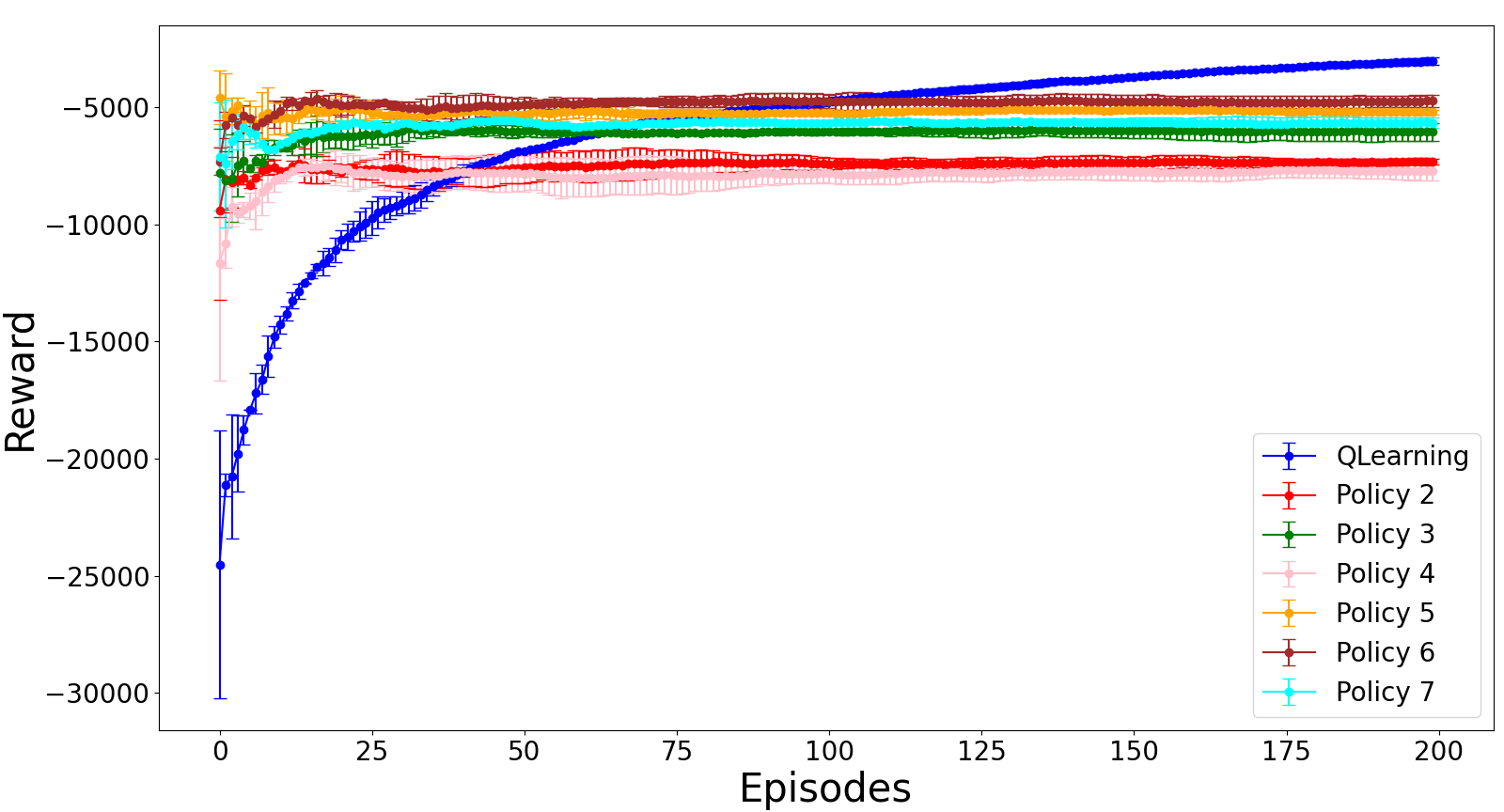}
\caption{Reuse Gain to learn market scenario 1 by reusing the policies learned in the other scenarios by $\pi$-reuse}
\label{fig:reuse-gain1}
\end{figure}

Finally, we use the $\pi$-reuse algorithm to compute the reuse gain, hence, the performance similarity between the market scenarios. Figure~\ref{fig:reuse-gain1} shows six learning processes, each one corresponding to the reuse of the policy learned in a source market (2-7) to learn in the target market 1 by applying $\pi$-reuse. Additionally, the blue line represents the average accumulated reward obtained by $Q$-learning. 

\begin{figure}[htbp]
\includegraphics[width=0.5\textwidth]{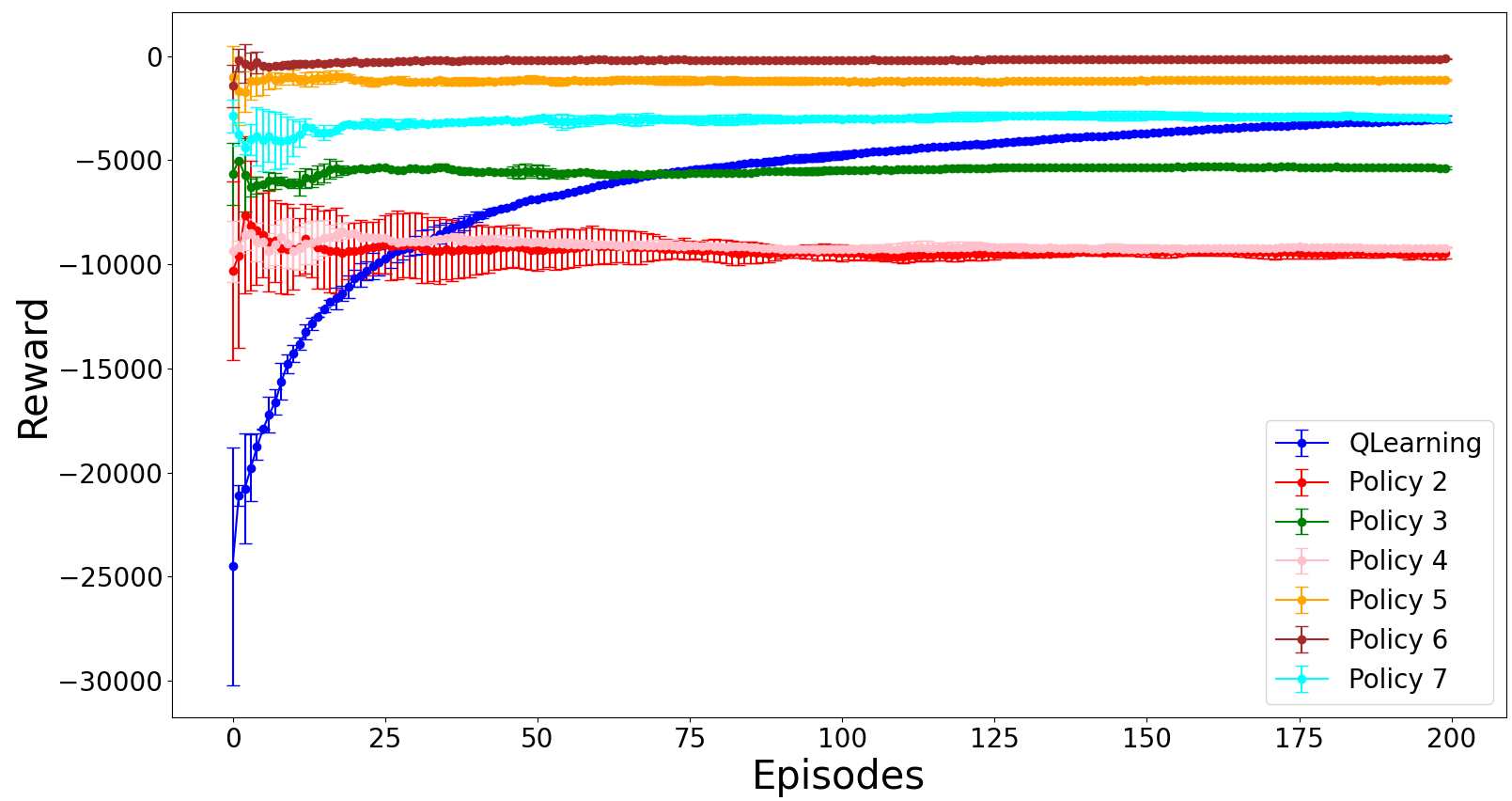}
\caption{Reuse Gain of the previously learned policies by $\pi$-reuse in a fully greedy setting.}
\label{fig:reuse-gain2}
\end{figure}

From Figure~\ref{fig:reuse-gain1}, it can be seen that $\pi$-reuse produces a jump-start at the initial episodes, although the final performance is lower than that obtained by $Q$-learning given that $\pi$-reuse always maintain active exploration. All policies achieve a similar reward, although the best seem to be policies 5 and 6. In Figure~\ref{fig:reuse-gain2}, the blue line represents the same as before, but the other lines represent the reuse gain obtained with the policies learned previously by $\pi$-reuse in a fully greedy setting. 

It can be observed that when the policies are followed fully greedy, the policy 5, 6 and 7, show a performance equal or superior to $Q$-learning, precisely because the exploration has been deactivated. Besides, it can be observed that policies 5 and 6, who are in the same cluster, are the best performers. 

Figures~\ref{fig:reuse-gain1} and~\ref{fig:reuse-gain2} show that there are two policies which obtain better results than the others. These policies are the the corresponding to the scenarios 5 and 6. Therefore, if we consider that more similar means higher final performance, then policies 5 and 6 are the most similar to scenario 1, the same conclusion we reached with the previous metrics.

\subsection{Transfer learning through PRQ-Learning}
\label{subsec:prq}

\begin{figure}[htbp]
\includegraphics[width=0.5\textwidth]{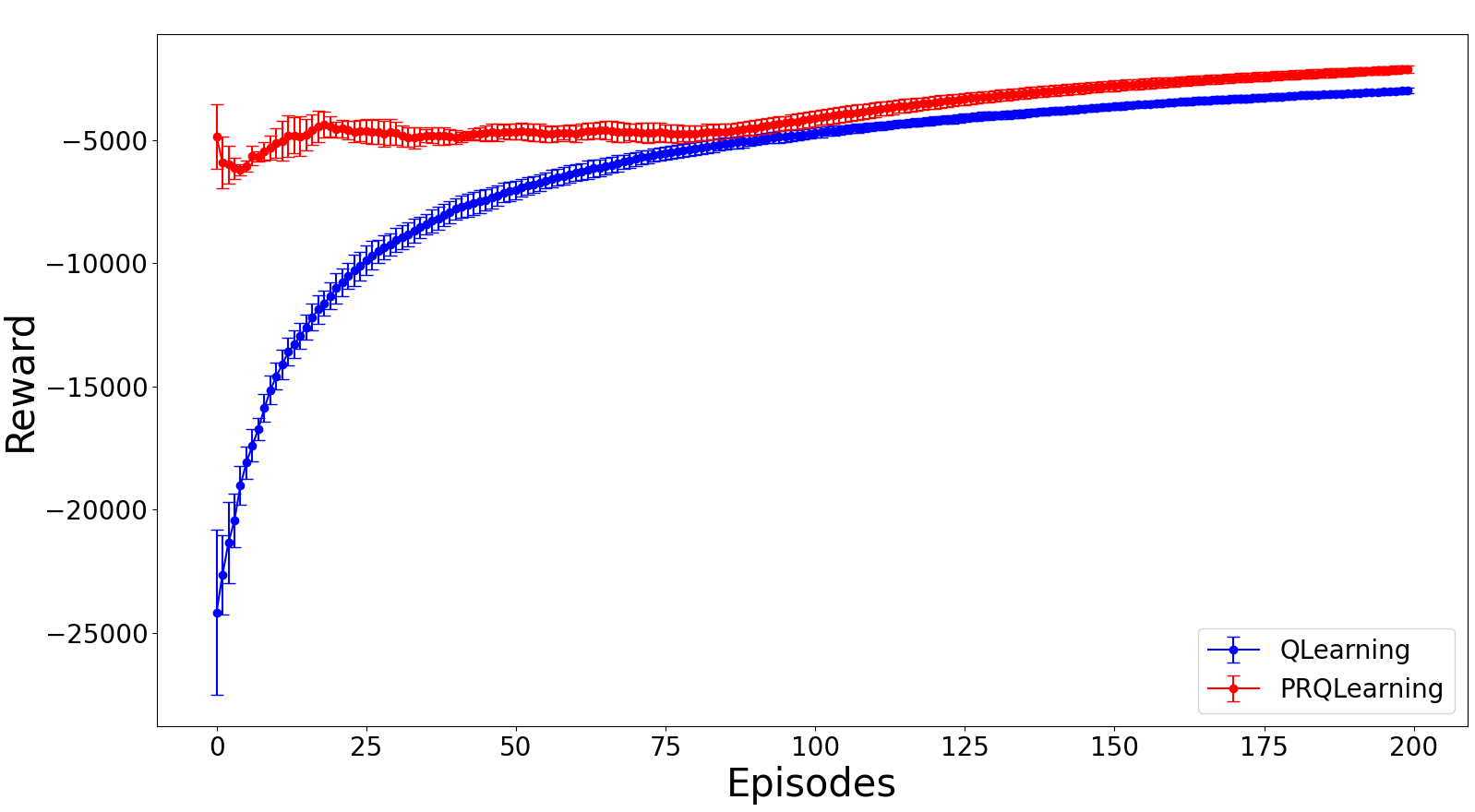}
\caption{Accumulated reward obtained by Q-learning vs PRQ-Learning}
\label{fig:prqVSql}
\end{figure}

Finally, the $Q$-Learner agent has been trained with PRQ-Learning in the scenario 1 reusing the policies learned in the scenarios 2-7 as a library. In Figure~\ref{fig:prqVSql} we compare the accumulated reward obtained with PRQ-Learning against $Q$-Learning.

\begin{figure}[htbp]
\includegraphics[width=0.5\textwidth]{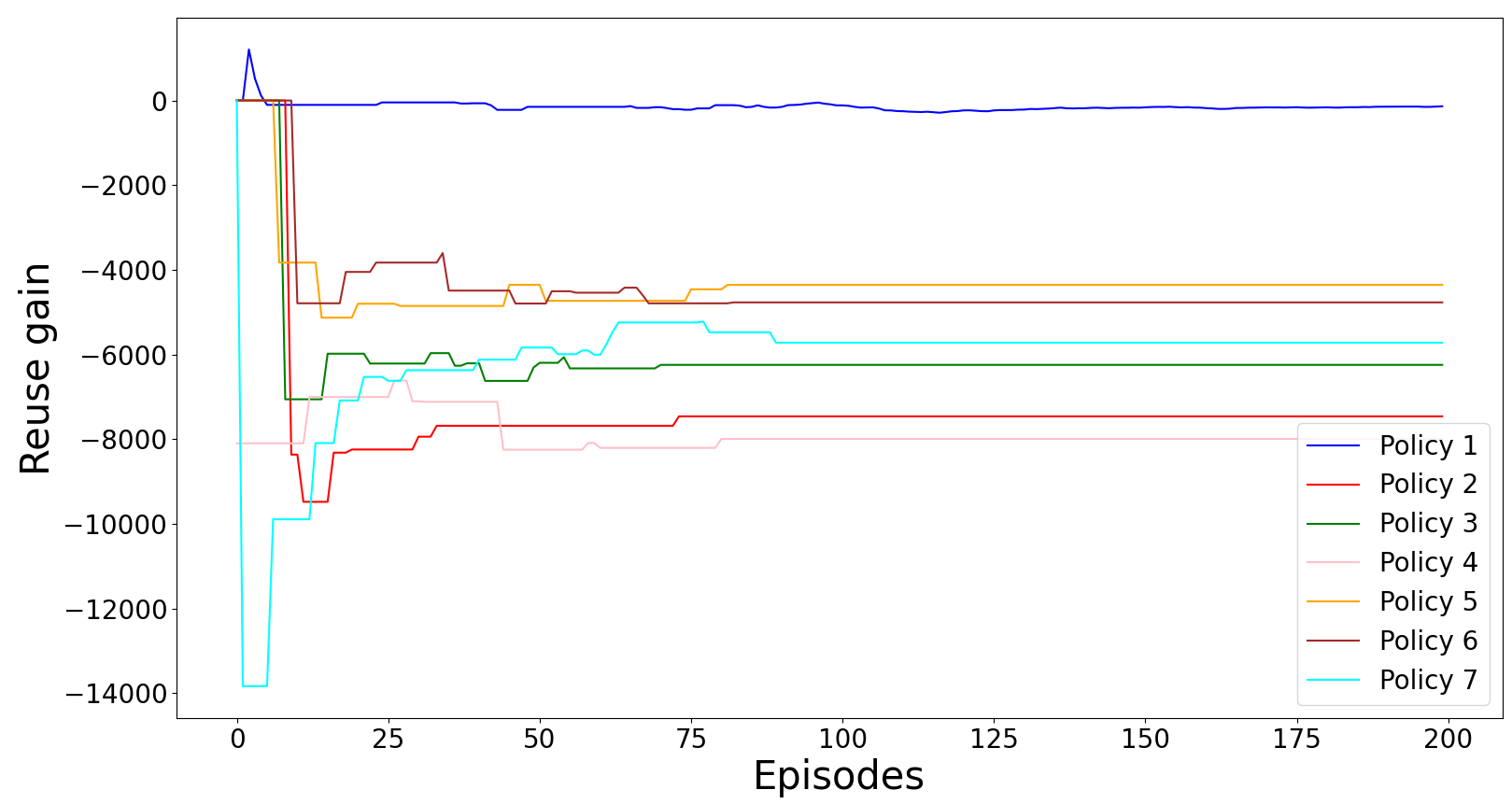}
\caption{Reuse gain evolution}
\label{fig:ReuseGain}
\end{figure}

Figure~\ref{fig:prqVSql} shows that PRQ-Learning achieves better results than $Q$-Learning in all episodes, also in the Figure~\ref{fig:ReuseGain} we can see that the two policies most similar to scenario 1 (policy 5 and 6), according the similarity metrics, are the ones which reach higher reuse gain.

\begin{figure}[htbp]
\includegraphics[width=0.5\textwidth]{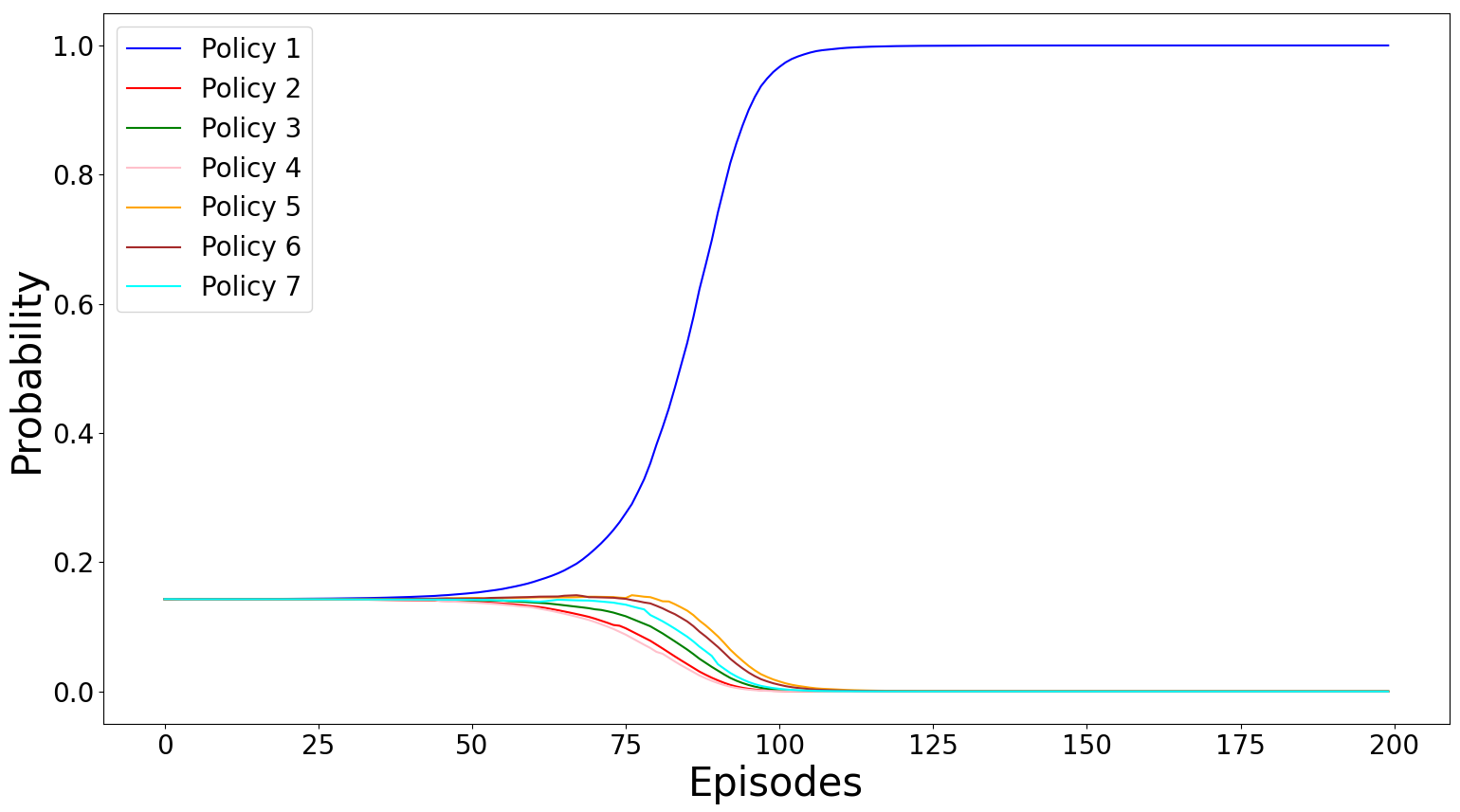}
\caption{Evolution of the probability of selecting each policy}
\label{fig:Probabilities}
\end{figure}

In Figure~\ref{fig:Probabilities} we can also see the probability of selecting each policy in each episode. It can be seen how at the beginning of the learning process all policies have the same probability of being selected. However, around episode 50 it can be seen how the probability of the current policy, together with policies 5 and 6 increases with respect to the others. Therefore, policies 5 and 6 have a higher probability of being selected because they are more similar with respect to scenario 1.

\section{Conclusions}
\label{sec:conclusions}

This paper has analyzed the use of three similarity metrics from a \textit{conceptual}, \textit{structural} and \textit{performance} perspective to measure the similarity between market scenarios. Analyzing the similarity between markets from different perspectives allows us to obtain a more complete analysis of what is similar and what is not. In the particular case of the market scenarios considered in this paper, the three similarity metrics considered have been able to determine which scenarios are more similar to each others, reaching the same conclusion and determining that there are two clusters according to whether the scenarios have momentum agents or not. Furthermore, through PRQ-Learning it is confirmed that the scenarios of the same cluster as the target scenario are the most similar to this one, and also the ones that improve learning the most. Therefore, by using similarity measures to determine which configurations are most similar to the target task and by using PRQ-Learning to transfer the knowledge learned in a test environment, learning in the target task could be improved. This is particular relevant in a \textit{Sim-to-Real} context: similarity metrics can help to answer how similar simulations and the actual world are. They could be used to provide theoretical guaranties that ensure the learned policies transferred from simulation to the actual world will perform as required, or to deﬁne mechanisms to tune/modify the simulated environments, so the gap between the simulated world and the actual one decreases. The latter opens new lines of research that we are currently investigating.

\section{Acknowledgements}

This research was funded in part by JPMorgan Chase Co. Any views or opinions expressed herein are solely those of the authors listed, and may differ from the views and opinions expressed by JPMorgan Chase Co. or its affiliates. This material is not a product of the Research Department of J.P. Morgan Securities LLC. This material should not be construed as an individual recommendation for any particular client and is not intended as a recommendation of particular securities, financial instruments or strategies for a particular client. This material does not constitute a solicitation or offer in any jurisdiction. This work has also been supported by the Madrid Government (Comunidad de Madrid-Spain) under the Multiannual Agreement with UC3M in the line of Excellence of University Professors (EPUC3M17), and in the context of the V PRICIT(Regional Programme of Research and Technological Innovation). Finally, Javier García is partially  supported by the Comunidad de Madrid funds under the project 2016-T2/TIC-1712.

\bibliography{main}

\begin{thebibliography}{28}
\providecommand{\natexlab}[1]{#1}
\providecommand{\url}[1]{\texttt{#1}}
\providecommand{\urlprefix}{URL }
\expandafter\ifx\csname urlstyle\endcsname\relax
  \providecommand{\doi}[1]{doi:\discretionary{}{}{}#1}\else
  \providecommand{\doi}{doi:\discretionary{}{}{}\begingroup
  \urlstyle{rm}\Url}\fi

\bibitem[{Ammar et~al.(2014)Ammar, Eaton, Taylor, Decebal, Mocanu, Driessens,
  Weiss, and Tuyls}]{inproceedings}
Ammar, H.; Eaton, E.; Taylor, M.; Decebal, C.; Mocanu, D.; Driessens, K.;
  Weiss, G.; and Tuyls, K. 2014.
\newblock An Automated Measure of MDP Similarity for Transfer in Reinforcement
  Learning.

\bibitem[{B{\"a}uerle and Rieder(2011)}]{bauerle2011markov}
B{\"a}uerle, N.; and Rieder, U. 2011.
\newblock \emph{Markov decision processes with applications to finance}.
\newblock Springer Science \& Business Media.

\bibitem[{Bouchaud et~al.(2018)Bouchaud, Bonart, Donier, and
  Gould}]{Bouchaud_book}
Bouchaud, J.-P.; Bonart, J.; Donier, J.; and Gould, M. 2018.
\newblock \emph{{Trades, quotes and prices: financial markets under the
  microscope}}.
\newblock Cambridge: Cambridge University Press.

\bibitem[{Byrd(2019)}]{byrd2019explaining}
Byrd, D. 2019.
\newblock Explaining Agent-Based Financial Market Simulation.
\newblock \emph{arXiv preprint arXiv:1909.11650} .

\bibitem[{Byrd, Hybinette, and Balch(2019)}]{byrd2019abides}
Byrd, D.; Hybinette, M.; and Balch, T.~H. 2019.
\newblock ABIDES: Towards High-Fidelity Market Simulation for AI Research.

\bibitem[{Chakraborty(2019)}]{chakraborty2019capturing}
Chakraborty, S. 2019.
\newblock Capturing financial markets to apply deep reinforcement learning.
\newblock \emph{arXiv preprint arXiv:1907.04373} .

\bibitem[{Fabretti(2013)}]{fabretti2013problem}
Fabretti, A. 2013.
\newblock On the problem of calibrating an agent based model for financial
  markets.
\newblock \emph{Journal of Economic Interaction and Coordination} 8(2):
  277--293.

\bibitem[{Fern{\'{a}}ndez and Veloso(2013)}]{Fernandez2013}
Fern{\'{a}}ndez, F.; and Veloso, M. 2013.
\newblock {Learning domain structure through probabilistic policy reuse in
  reinforcement learning}.
\newblock \emph{Progress in Artificial Intelligence} 2(1): 13--27.
\newblock ISSN 21926360.

\bibitem[{Ganesh et~al.(2019)Ganesh, Vadori, Xu, Zheng, Reddy, and
  Veloso}]{ganesh2019reinforcement}
Ganesh, S.; Vadori, N.; Xu, M.; Zheng, H.; Reddy, P.; and Veloso, M. 2019.
\newblock Reinforcement Learning for Market Making in a Multi-agent Dealer
  Market.

\bibitem[{Huang(2018)}]{huang2018financial}
Huang, C.~Y. 2018.
\newblock Financial trading as a game: A deep reinforcement learning approach.
\newblock \emph{arXiv preprint arXiv:1807.02787} .

\bibitem[{Kaelbling, Littman, and Moore(1996)}]{Kaelbling1996ReinforcementLA}
Kaelbling, L.; Littman, M.; and Moore, A. 1996.
\newblock Reinforcement Learning: A Survey.
\newblock \emph{J. Artif. Intell. Res.} 4: 237--285.

\bibitem[{Kanwar et~al.(2019)}]{kanwar2019deep}
Kanwar, N.; et~al. 2019.
\newblock \emph{Deep Reinforcement Learning-Based Portfolio Management}.
\newblock Ph.D. thesis.

\bibitem[{Kyle(1985)}]{kyle1985continuous}
Kyle, A.~S. 1985.
\newblock Continuous auctions and insider trading.
\newblock \emph{Econometrica: Journal of the Econometric Society} 1315--1335.

\bibitem[{Mahmud et~al.(2013)Mahmud, Hawasly, Rosman, and
  Ramamoorthy}]{MahmudM.M.Hassan2013CMDP}
Mahmud, M. M.~H.; Hawasly, M.; Rosman, B.; and Ramamoorthy, S. 2013.
\newblock {Clustering Markov Decision Processes For Continual Transfer} .

\bibitem[{Mart{\'i}nez, Garc{\'i}a, and
  Fern{\'a}ndez(2020)}]{martinez2020probabilistic}
Mart{\'i}nez, E.; Garc{\'i}a, J.; and Fern{\'a}ndez, F. 2020.
\newblock Probabilistic Policy Reuse for Similarity Computation Among Market
  Scenarios.
\newblock \emph{FinPlan 2020} 28.

\bibitem[{Mehta et~al.(2008)Mehta, Natarajan, Tadepalli, and Fern}]{Mehta2008}
Mehta, N.; Natarajan, S.; Tadepalli, P.; and Fern, A. 2008.
\newblock {Transfer in variable-reward hierarchical reinforcement learning}.
\newblock \emph{Machine Learning} 73(3): 289--312.
\newblock ISSN 08856125.

\bibitem[{Moody et~al.(1998)Moody, Wu, Liao, and
  Saffell}]{Performancefunctions}
Moody, J.; Wu, L.; Liao, Y.; and Saffell, M. 1998.
\newblock Performance functions and reinforcement learning for trading systems
  and portfolios.
\newblock \emph{Journal of Forecasting} 17(5‐6): 441--470.

\bibitem[{Nevmyvaka, Feng, and Kearns(2006)}]{nevmyvaka2006reinforcement}
Nevmyvaka, Y.; Feng, Y.; and Kearns, M. 2006.
\newblock Reinforcement learning for optimized trade execution.
\newblock In \emph{Proceedings of the 23rd international conference on Machine
  learning}, 673--680.

\bibitem[{Onta{\~n}{\'o}n(2020)}]{ontanon2020overview}
Onta{\~n}{\'o}n, S. 2020.
\newblock An overview of distance and similarity functions for structured data.
\newblock \emph{Artificial Intelligence Review} 53(7): 5309--5351.

\bibitem[{R.Cont(2001)}]{RCont2001}
R.Cont. 2001.
\newblock Empirical properties of asset returns: stylized facts and statistical
  issues.
\newblock \emph{Quantitative Finance} 1(2): 223--236.
\newblock \doi{10.1080/713665670}.

\bibitem[{Salakhutdinov, Mnih, and Hinton(2007)}]{RBMS}
Salakhutdinov, R.; Mnih, A.; and Hinton, G. 2007.
\newblock Restricted Boltzmann Machines for Collaborative Filtering.
\newblock In \emph{Proceedings of the 24th International Conference on Machine
  Learning}, ICML '07, 791–798. New York, NY, USA: Association for Computing
  Machinery.
\newblock ISBN 9781595937933.
\newblock \doi{10.1145/1273496.1273596}.
\newblock \urlprefix\url{https://doi.org/10.1145/1273496.1273596}.

\bibitem[{Sunmola and Wyatt(2006)}]{Sunmola2006}
Sunmola, F.~T.; and Wyatt, J.~L. 2006.
\newblock {Model transfer for Markov decision tasks via parameter matching}.
\newblock \emph{In Proceedings of the 25th Workshop of the UK Planning and
  Scheduling Special Interest Group (PlanSIG 2006)} .

\bibitem[{Sutton and Barto(2018)}]{sutton2018reinforcement}
Sutton, R.~S.; and Barto, A.~G. 2018.
\newblock \emph{Reinforcement learning: An introduction}.
\newblock MIT press.

\bibitem[{Tan et~al.(2018)Tan, Zhang, Coumans, Iscen, Bai, Hafner, Bohez, and
  Vanhoucke}]{tan2018sim}
Tan, J.; Zhang, T.; Coumans, E.; Iscen, A.; Bai, Y.; Hafner, D.; Bohez, S.; and
  Vanhoucke, V. 2018.
\newblock Sim-to-real: Learning agile locomotion for quadruped robots.
\newblock \emph{arXiv preprint arXiv:1804.10332} .

\bibitem[{Taylor and Stone(2009)}]{taylor2009transfer}
Taylor, M.~E.; and Stone, P. 2009.
\newblock Transfer learning for reinforcement learning domains: A survey.
\newblock \emph{Journal of Machine Learning Research} 10(7).

\bibitem[{Torrey and Shavlik(2010)}]{torrey2010transfer}
Torrey, L.; and Shavlik, J. 2010.
\newblock Transfer learning.
\newblock In \emph{Handbook of research on machine learning applications and
  trends: algorithms, methods, and techniques}, 242--264. IGI global.

\bibitem[{Vyetrenko et~al.(2020)Vyetrenko, Byrd, Petosa, Mahfouz, Dervovic,
  Veloso, and Balch}]{vyetrenko2019real}
Vyetrenko, S.; Byrd, D.; Petosa, N.; Mahfouz, M.; Dervovic, D.; Veloso, M.; and
  Balch, T.~H. 2020.
\newblock Get Real: Realism Metrics for Robust Limit Order Book Market
  Simulations.
\newblock \emph{International Conference on AI in Finance.} .

\bibitem[{Watkins(1989)}]{watkins1989}
Watkins, C. 1989.
\newblock \emph{Learning from Delayed Rewards}.
\newblock Ph.D. thesis, King's College, Cambridge, UK.

\end{thebibliography}

\end{document}